\documentclass[preprint]{aastex631}
\usepackage{natbib}
\accepted{4 December 2023}

\submitjournal{ApJ}

\begin{document}


\newcommand{\ltsimeq}{\raisebox{-0.6ex}{$\,\stackrel{\raisebox{-.2ex}%
{$\textstyle<$}}{\sim}\,$}}
%
\newcommand{\gtsimeq}{\raisebox{-0.6ex}{$\,\stackrel{\raisebox{-.2ex}%
{$\textstyle>$}}{\sim}\,$}}


\title{X-ray Observations of the Enigmatic Wolf-Rayet System $\theta$ Mus: \\
       Two's Company But Three's a Crowd}

\correspondingauthor{Stephen L. Skinner}
\email{stephen.skinner@colorado.edu}

\author{Stephen L. Skinner}
\affiliation{Center for Astrophysics and
Space Astronomy (CASA), University of Colorado,
Boulder, CO, USA 80309-0389}

\author{Svetozar A. Zhekov}
\affiliation{Inst. of Astronomy and National Astronomy Observatory,
Bulgarian Academy of Sciences,
72 Tsarigradsko Chaussee Blvd., Sofia, 1784, Bulgaria}

\author{Manuel G\"{u}del}
\affiliation{Dept. of Astrophysics, Univ. of Vienna,
T\"{u}rkenschanzstr. 17,  A-1180 Vienna, Austria}

\author{Werner Schmutz}
\affiliation{Physikalisch-Meteorologisches Observatorium Davos and
World Radiation Center (PMOD/WRC), Dorfstrasse 33, CH-7260 Davos Dorf, Switzerland}

\begin{abstract}
 $\theta$ Mus is a remarkable spectroscopic binary (SB) consisting of a 
 carbon-type Wolf-Rayet star and OV companion (WC6$+$O6-7V) in a 
 $\approx$19-day orbit. In addition an O-supergiant is visually detected 
 at a small offset of 46 mas and if gravitationally bound to the SB system would have
 an orbital period of many decades. $\theta$ Mus is  X-ray bright and a nonthermal 
 radio source as commonly observed in massive colliding wind (CW) binaries. 
 We present new {\em Chandra} X-ray observations of $\theta$ Mus which
 complement previous {\em XMM-Newton} observations. The X-ray emission 
 consists of a cool nearly steady weakly-absorbed plasma component with broad redshifted
 emission lines located in an extended region far from the SB system. 
 Hotter plasma is also present traced by Fe XXV emission. The observed flux
 in the $\approx$2-5 keV range dropped significantly on a timescale of $\leq$5 years.
 The flux decrease can be attributed to an increase in absorption toward the
 hotter plasma which is likely located in the confined wind interaction region
 of the short-period SB system. The X-ray emission of $\theta$ Mus is remarkably
 similar to the WC$+$O binary $\gamma^2$ Vel including carbon recombination spectral 
 lines but both systems show unusual line centroid properties that challenge CW models.
\end{abstract}


\vspace*{3.0cm}

\section{Introduction}
Wolf-Rayet (WR) stars are evolved massive stars that may end
their lives as supernovae, in some cases accompanied by 
$\gamma$-ray bursts (Woosley \& Bloom 2006).
They are losing mass at high rates and their powerful winds 
replenish the interstellar medium with metal-rich material 
that will be recycled into new generations of stars.
With typical terminal
wind speeds V$_{\infty}$ $\approx$ 1000 - 2500 km s$^{-1}$, 
the supersonic winds of WR stars and interacting winds in massive 
WR$+$OB binaries offer excellent opportunities for observational
tests of wind shock models.

The present work focuses on X-ray properties of the multiple
system $\theta$ Mus (= HD 113904 = WR 48). It is a 
spectroscopic binary (SB) consisting of a carbon-type WR star
with an O-star companion (WC6$+$O6-7V) in a 19-day orbit.
In addition an O9.5I star is visually detected
at an offset of 46$\pm$9 mas (Hartkopf et al. 1999).
{\em Chandra} has an angular resolution of about one arcsecond which is not
sufficient to spatially resolve the two components of the $\theta$ Mus SB system
or disentangle emission of the SB system from that of the O supergiant.
X-rays in such multiple systems can arise in the shocked
winds of either star or in colliding wind (CW) shocks formed 
between the stars. Thus, the observed X-ray spectrum
may consist of multiple superimposed components. Distinguishing
between them relies on diagnostics such as plasma temperature 
distribution and emission line properties (e.g. line widths, 
centroid shifts). Reliable measurements of line properties
require high spectral resolution grating observations.

X-rays can be produced in the winds of the individual stars in massive
binaries as a result of radiative shocks associated with line-driven 
instabilities (Lucy \& White 1980; Owocki et al. 1988; Feldmeier et al. 1997).
Radiative shocks are predicted to generate cool X-ray plasma
(kT $\ltsimeq$ 1 keV) as has been observed in objects 
like the O4I star $\zeta$ Puppis (Cassinelli et al. 2001; Kahn et al. 2001). 
But single nitrogen-type WN stars without known companions
reveal hotter plasma (kT $\gtsimeq$ 2 keV) that is not predicted for 
radiative wind shocks (Skinner et al. 2010; 2012) as do rare oxygen-type
WO stars (Sokal et al. 2010; Skinner et al. 2019).
But it is remarkable  that putatively
single carbon-type WC stars with tremendous bolometric
luminosities and powerful winds have so far eluded X-ray detection 
in sensitive pointed observations (Oskinova et al. 2003;  Skinner et al. 2006).
 
Hot X-ray plasma can arise in CW shocks in WR$+$OB systems 
(Cherapashchuk 1976; Prilutskii \& Usov 1976; Luo et al. 1990; Usov 1992;
Stevens, Blondin, \& Pollock 1992). 
The predicted maximum temperature for an adiabatic CW shock is
kT$_{cw}$ $\approx$ 1.96$\mu$[V$_{\perp}$/1000 km s$^{-1}$]$^{2}$~keV
where $\mu$ is the mean  particle weight in
the wind and V$_{\perp}$ is the wind velocity component perpendicular
to the shock front (Luo et al. 1990). For terminal wind speeds 
V$_{\infty}$ $\approx$ 1000 - 2500 km s$^{-1}$ typical of WR and O stars
maximum shock temperatures on the line-of-centers 
kT$_{cw}$  $\gtsimeq$ 2 keV are expected. Such hotter plasma has
now been detected in several WR binaries. These include 
well-studied systems like the 78-day binary $\gamma^2$ Vel whose
whose WC8$+$O7.5 type is similar to $\theta$ Mus 
(Willis, Schild, \& Stevens 1995; Skinner et al. 2001; Schild et al. 2004).

We  present here new {\em Chandra} X-ray observations of
the multi-component WR system $\theta$ Mus when the O6-7V
star was passing in front of the WR star ($\phi$ $\approx$ 0.6).
In addition we summarize archived {\em XMM-Newton} data 
obtained in 2004 at nearly the same phase as {\em Chandra} 
and in 2009 ($\phi$ $\approx$ 0.95) with the WR star passing
in front. The primary objective was to confirm and extend 
previous reports of redshifted emission lines and
attempt to reconcile the observed X-ray emission with 
the CW shock interpretation.

\clearpage

\section{The Target: $\theta$ Mus}

$\theta$ Mus (= HD 113904) is listed as WR 48 in the catalog of
van der Hucht (2001) with a visual extinction A$_{v}$ = 0.93 mag.
This equates to A$_{V}$ = 0.84 mag using A$_{V}$ = 0.9A$_{v}$. 
Its {\em Gaia} DR3 parallax distance is 2.168 kpc. The SB system was classified as 
WC6$+$O6-7V by Hill et al. (2002).

To determine orbital phases of the X-ray observations we 
use the orbital solution for the SB system 
of Hill et al. (2002) who adopted a period 
P$_{orb}$ = 19.1375$\pm$0.0025 d based on the work of 
Schnurr (1999).  But the determination of the SB orbital 
parameters is hampered by the nearby bright supergiant. 
No atmospheric eclipse that could be used
to constrain the orbital ephemeris has so far been
detected in the SB system (Lenoir-Craig et al. 2021).
Lenoir-Craig et al. also note an unexplained phase difference
for time of minimum brightness between their results
and those of Marchenko et al. (1998). Attempts to
quantify the SB orbit based on linear polarization
observations have yielded negative results (St.-Louis et al. 1987).
Moffat \& Seggewiss (1977) determined P$_{orb}$ = 18.341$\pm$0.008 d,
but aliasing also allows values of 18.596 d, 18.858 d, or 19.128 d
(Lenoir-Craig et al. 2021). Using {\em Hipparcos} photometry
Marchenko et al. (1998) obtained P$_{orb}$ = 18.05$\pm$0.32 d.
Considering the above mixed results, the X-ray orbital phases
computed using P$_{orb}$ = 19.1375 d (Tables 1,2) should
be treated with caution.

If the OI star is 
gravitationally bound to the SB system at the {\em Gaia}
DR3 parallax distance of 2.17 kpc then a long-period orbit is predicted.
The 46 mas offset corresponds to a {\em projected} 
separation $D_{proj}$ $\approx$ 100 AU. To estimate the orbital period
we assume a WR mass
M$_{WC6}$ = 12 M$_{\odot}$ (Hill et al. 2002) and 
O-star masses (Weidner \& Vink 2010) 
M$_{O7V}$ = 30 M$_{\odot}$ and M$_{O9.5I}$ = 35 M$_{\odot}$.
To simplify the three-body orbit to a tractable two-body
orbit we combine the SB system into a single object of
mass $\approx$42 M$_{\odot}$. This combined mass is similar
to the OI star mass and the center-of-mass thus lies 
approximately halfway between the SB system and the OI star. 
If the orbit of the OI$+$SB system is near-circular
their physical (deprojected) separation $D$ is twice the orbit 
semi-major axis $a$.
Since $D$ $\geq$ $D_{proj}$ we obtain $a$ $\geq$ 50 AU
and by Kepler's 3rd law  P$_{orb,OI}$ $\geq$40 years. 
This is a lower limit and the actual period, which 
depends on the (unknown) {\em deprojected} separation $D$,
could be much longer than 40 years.
Previous nonthermal radio studies have argued 
that the OI star is in such a bound orbit 
(Dougherty \& Williams 2000).
 
\section{Previous {\em XMM-Newton} Observations}

An X-ray observation of $\theta$ Mus with {\em XMM-Newton}
was obtained in 2004 (ObsId 0090030201) when the OV star was nearly 
in front of the WR star and 
analyzed by Sugawara et al. (2008). A second {\em XMM-Newton} 
observation in 2009 (ObsId 0605670201) at nearly opposite phase 
is archived and discussed below.
Somewhat surprisingly, bright low-temperature 
emission lines such as O VIII (maximum line power at T$_{max}$ $\approx$ 3 MK)
in the {\em XMM-Newton}  grating spectrum 
were found to be  redshifted by  $\approx$650 km s$^{-1}$ when the 
O6-7V star was passing in front. 
If the X-ray emission originates in a CW shock in the SB system and 
the WR wind momentum dominates the O6-7V wind (as is usually the case)
then blueshifted lines would be expected when the O6-7V star is passing 
in front of the WR star. In that viewing geometry the CW shock cone wraps 
around the  O6-7V star with a blueshifted flow velocity component 
toward the observer. The magnitude of the blueshift depends on the 
shock cone opening angle with wider angles giving smaller blueshifts.
To explain the puzzling redshifted lines, 
Sugawara et al. (2008) proposed that they originate in a CW shock 
between the SB system and the OI star. In that picture 
the OI star would be in a bound orbit with the SB system and 
behind it (Fig. 4 of Sugawara et al. 2008).

\begin{deluxetable}{lcccccccc}
\tabletypesize{\scriptsize}
\tablewidth{0pt}
\tablecaption{$\theta$ Mus X-ray Properties (Chandra ACIS-S/HETG 0-order)}
\tablehead{
           \colhead{ObsId}               &
           \colhead{Start Date/Time}     &
           \colhead{Livetime}            &
           \colhead{Phase\tablenotemark{\scriptsize a}}               &
           \colhead{Counts}              &
           \colhead{Count rate}          &
           \colhead{E$_{\rm 50}$}        &
           \colhead{H.R.}                &
           \colhead{P$_{var}$}           \\
           \colhead{}                    &
           \colhead{TT}                  &
           \colhead{(ks)}                &
           \colhead{}                    &
           \colhead{(cts)}                 &
           \colhead{(cts ks$^{-1}$)}       &
           \colhead{(keV)}               &
           \colhead{}                    &
           \colhead{}                  
}
\startdata
23378        & 2020 Nov 24~10:39   & 14.757           &  0.60-0.61  & 189 & 12.56$\pm$2.98 & 2.60        & 0.67        & 0.06  \\
24869        & 2020 Nov 24~23:16   & 14.755           &  0.63-0.64  & 211 & 13.91$\pm$3.11 & 2.83        & 0.79        & 0.12  \\
24497        & 2022 Jun 1 ~19:25   & 29.574           &  0.57-0.59  & 259 &  8.63$\pm$2.99 & 2.79        & 0.71        & 0.09  \\
24498        & 2022 Jun 3 ~06:03   & 30.393           &  0.64-0.66  & 408 & 13.17$\pm$2.48 & 2.64        & 0.74        & 0.06  \\
24496        & 2022 Jun 22~13:14   & 29.575           &  0.65-0.67  & 345 & 11.60$\pm$3.71 & 2.63        & 0.74        & 0.05  \\
mean         &  ...                & ...              &  ...        & ... & 11.97$\pm$1.83 & 2.70        & 0.73        & ...   \\
\enddata
\tablecomments{
The livetime is the on-source time and excludes dead-time when the detector is not collecting data.
X-ray properties are based on source events in the 0.2 - 8 keV range.
The mean count rates and standard deviations were obtained by fitting the X-ray light curve 
binned at 1200 s intervals. E$_{\rm 50}$ is the median photon energy. 
Hardness ratio H.R. = counts(2-8 keV)/counts(0.2-8 keV).
P$_{var}$ is the probability that the source is variable based on event arrival times  
as determined by the CIAO tool {\em glvary}. 
The X-ray centroid position obtained by averaging all observations is
(J2000) R.A. = 13h 08m 07.140s Decl. = $-$65$^{\circ}$ 18$'$ 21.70$''$.
The {\em Hubble Space Telescope} ({\em HST}) Guide Star Catalog (GSC v2.3.2) 
position of $\theta$ Mus
is R.A. = 13h 08m 07.155s Decl. = $-$65$^{\circ}$ 18$'$ 21.50$''$.
The 2MASS position is
 R.A. = 13h 08m 07.153s Decl. = $-$65$^{\circ}$ 18$'$ 21.517$''$.
\tablenotetext{a}{The start-stop phases are referenced to phase $\phi$ = 0 (WR star in front) at
T$_{o}$ = HJD 2451377.51$\pm$0.45 d and assume an orbital period P$_{orb}$= 19.1375$\pm$0.0025 d (Hill et al. 2002).}
}
\end{deluxetable}

\section{X-ray Observations}

\subsection{Chandra}

The total {\em Chandra} observing time ($\approx$120 ks) 
was split into five segments (Table 1) to  accommodate {\em Chandra}'s
operational constraints. 
The data were obtained using the High Energy Transmission
Grating (HETG)/Advanced CCD Imaging Spectrometer (ACIS-S) combination.
The HETG provides grating data from the Medium Energy Grating (MEG)
and High Energy Grating (HEG). ACIS-S also gives undispersed HETG 0-order
data (hereafter referred to as ACIS-S 0-order or simply ACIS-S) at lower
energy resolution than HETG. Detailed information on {\em Chandra}'s instrumentation
is in the Proposer's Observatory Guide
(POG)\footnote{https://cxc.harvard.edu/proposer/POG/}.

The pipeline-processed data files provided by the {\em Chandra} X-ray
Center were analyzed using science
threads in {\em Chandra} Interactive Analysis Software (CIAO version 4.14)
and recent calibration data (CALDB version 4.9.8). 
ACIS-S 0-order source events were extracted from a circular region of
radius 2.$''$0 centered on the X-ray peak for each observation.
Background is negligible. Energy-filtered 0-order source events were 
used to determine count rates, median event energies, and hardness ratios (H.R.). 
The  probability of source variability was determined using the 
CIAO tool {\em glvary}. CIAO {\em specextract} was used to extract ACIS-S 0-order
spectra for each observation along with observation-specific response matrix 
files (RMFs) and auxiliary response files (ARFs). Spectral fitting was 
undertaken with  CIAO {\em Sherpa} and HEASOFT XSPEC 
version 12.8.2\footnote{http://heasarc.gsfc.nasa.gov/xanadu/xspec}.
The ACIS-S 0-order spectra for the five observations were fitted 
simultaneously in XSPEC, the recommended procedure for analyzing 
{\em Chandra} data acquired at different epochs.
The five spectra were combined into a single spectrum
for display purposes. To facilitate comparison with the results of
Sugawara et al. (2008) we have modeled X-ray absorption 
using the  XSPEC $wabs$ model. Fits were compared with the more recent
$tbabs$ absorption model and only minor differences were found.

We analyzed 1st order MEG1 and HEG1 grating spectra ($+$1 and $-1$ orders combined).
The MEG1 spectra for each observation were combined into a summed MEG1 spectrum
using CIAO {\em combine\_grating\_spectra}, and similarly for HEG1.
Grating RMF and ARF files were generated using CIAO {\em mktgresp}. 
Emission lines were fitted with Gaussian profiles to determine line properties.
Since the grating spectra of the individual observations do not contain 
enough line counts to reliably measure line properties the summed
grating spectra were used for line fitting.

\subsection{XMM-Newton}

For comparison with {\em Chandra} we have retrieved the {\em XMM-Newton}
archive data for the two $\theta$ Mus observations, ObsIds 0090030201 
(119.6 ks, 20 July 2004) and 0605670201 (54.3 ks, 19 July 2009). The data 
were reprocessed using the {\em XMM-Newton}
Science Analysis Software (SAS vers. 18) modules {\em epchain} and  {\em emchain},
applying recent calibration data.
Undispersed source spectra were extracted using  SAS 
{\em xmmselect} for the pn, MOS1, and MOS2 
detectors comprising the European Photon Imaging Camera (EPIC).   
Background spectra were extracted in adjacent source free regions.
Observation-specific RMF and ARF files were created for each EPIC spectrum
using SAS modules {\em rmfgen} and {\em arfgen}. 
Spectra were fitted using XSPEC as for the {\em Chandra} data.  

\begin{figure}
\figurenum{1}
\includegraphics*[width=7.0cm,angle=0]{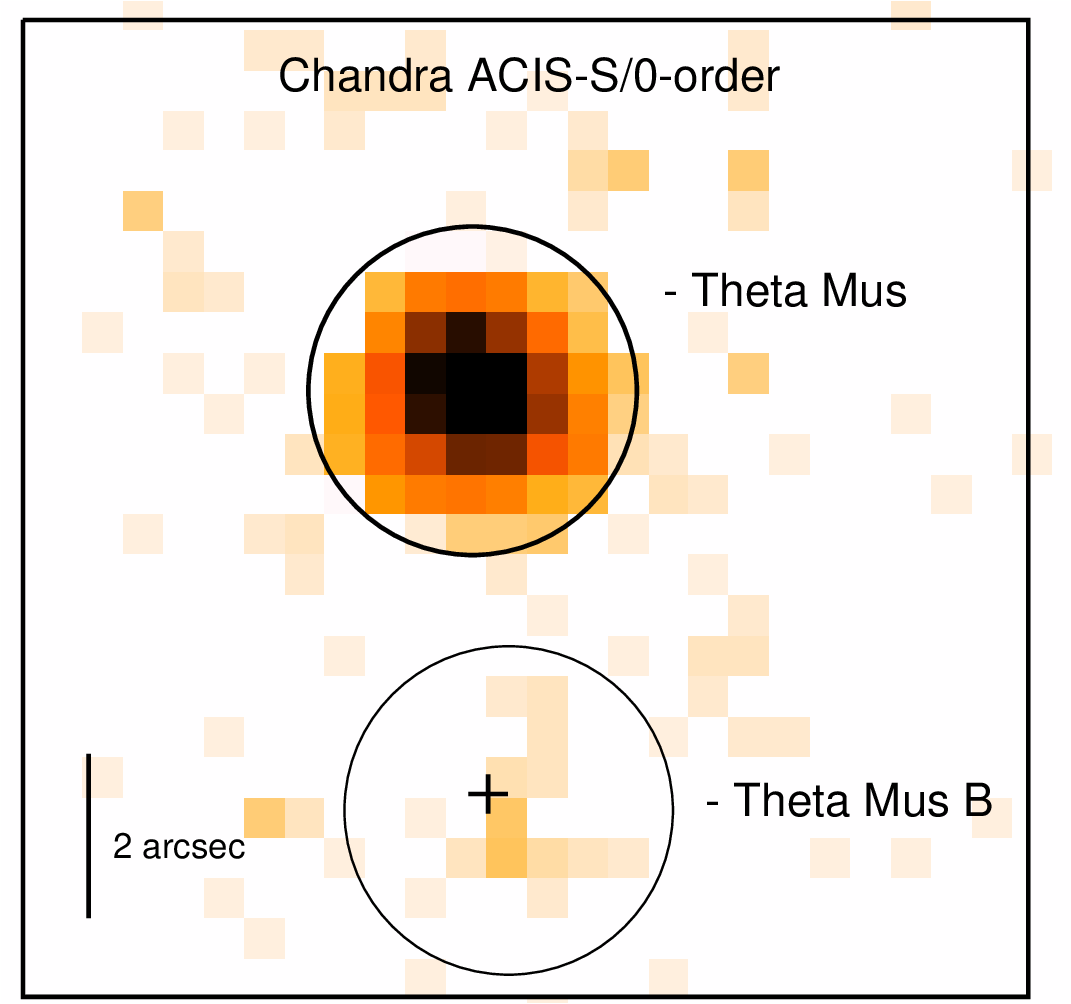}
\caption{{\em Chandra} image from summed June 2022 observations 
(0.2-8 keV; 0$''$.49 pixels; 89.54 ks livetime).
Circles (2$''$ radii) are centered on X-ray centroids. Cross ($+$) marks the 
optical position of $\theta$ Mus B which is likely
the faint X-ray source (16 cts).  N up, E left.}
\end{figure}

\section{Results}

\subsection{Chandra}

$\theta$ Mus was clearly detected in all five {\em Chandra} 
observations (Table 1).
Figure 1 shows the ACIS-S 0-order image obtained by combining
events from the June 2022 observations. A faint X-ray source
lies  5$''$.2 south of $\theta$ Mus whose X-ray
centroid is offset 0$''$.35 from the O9III star $\theta$ Mus B.
This offset is well within the 1$\sigma$ {\em Chandra} ACIS-S absolute 
astrometric accuracy of 0$''$.79 (68\% confidence 
circle radius\footnote{https://cxc.harvard.edu/cal/ASPECT/celmon/}).
so the faint source is most likely $\theta$ Mus B. Its
count rate is more than 60 times lower than $\theta$ Mus.  
The $r$ = 2$''$ extraction region used for  
$\theta$ Mus ACIS-S analysis excludes the faint emission 
from $\theta$ Mus B which is not of primary interest in this study.  
The X-ray centroid is in excellent agreement with optical and
near-IR positions of $\theta$ Mus. The X-ray properties 
in Table 1 are similar for each observation. The individual
observations show no significant X-ray variability but the
count rate in ObsId 24497 is 1.8$\sigma$ lower than the mean for all
five observations. If real, this decrease would indicate
variability on a short timescale of $\leq$2 days.

{\em Undispersed ACIS-S Spectra}.~ 
The summed ACIS-S spectrum is shown in Figure 2-top.
As noted, the individual spectra were fitted simultaneously
in XSPEC. We compared fits using two-temperature optically thin 
plasma models  (2T $vapec$) and
2T plane-parallel shock models (2T $vpshock$) which are 
typically used for modeling WR star X-ray spectra.
The $vapec$ model assumes collisional ionization equilibrium (CIE)
whereas the $vpshock$ model accounts for possible
non-equilibrium ionization (NEI) conditions that may
occur in the plasma of rapidly-heated shocks (e.g. Zhekov 2007).

To make comparisons  with the analysis of
{\em XMM-Newton} spectra by Sugawara et al. (2008)
we referenced the abundances in spectral fits (Table 2)
to the solar values of Anders \& Grevesse (1989). 
But evolved WC stars have nonsolar
abundances that are  depleted in hydrogen and
nitrogen with enhanced carbon. We have thus compared solar
abundance spectral fits with fits using generic WC star
abundances (van der Hucht, Cassinelli, \& Williams 1986).
The fit quality as judged by $\chi^2$ is poor using 
the generic WC abundances but improves
by allowing the C abundance to increase above the 
generic reference value, as was also true 
for fits referenced to solar abundances.

The 2T $vpshock$ model provides slightly better fits of the 
ACIS-S spectra than 2T $vapec$  as gauged by the 
$\chi^2$ fit statistic. Thus we do not include the ACIS-S 2T $vapec$ fit 
in Table 2 but it is included for the higher signal-to-noise ratio
2009 {\em XMM-Newton} EPIC data (discussed below). 
Even so, both models converge to similar values for physical 
quantities such as plasma temperature.
The ACIS-S spectra do not tightly constrain the absorption column
density so it was held fixed in ACIS-S fits at
N$_{\rm H}$ = 1.6 $\times$10$^{21}$ cm$^{-2}$, consistent with 
that expected for A$_{V}$ = 0.84 and a standard conversion
N$_{\rm H}$ (cm$^{-2}$) = 1.9$\pm$0.3 $\times$10$^{21}$$\cdot$A$_{\rm V}$
(Gorenstein 1975; Vuong et al. 2003). It is also nearly identical to 
that obtained from 2T $vpshock$ fits of {\em XMM-Newton} EPIC spectra (Table 2).

The various ACIS-S spectral fits show very little
sensitivity to abundance deviations from solar, which is not
surprising given the moderate spectral resolution of ACIS-S
and the modest number of spectral counts.
However all fits converged to a subsolar Fe abundance,
as was also found in our analysis of {\em XMM-Newton}
spectra below and by  Sugawara et al. (2008).
The carbon and oxygen lines present in {\em XMM-Newton} spectra at energies 
below 0.7 keV lie below the energy range where ACIS-S is sensitive.
Thus {\em Chandra}  does not provide any constraints on 
the C abundance although {\em XMM-Newton} spectra do (Sec. 5.2).

{\em HETG Spectra}.~
The {\em Chandra} MEG1 spectrum of $\theta$ Mus is shown in Figure 3-top.
Several emission lines and line blends are present (Table 3) spanning a 
temperature range of T$_{max}$ $\approx$ 6 MK (Ne X) to
T$_{max}$ $\approx$ 16 MK (S XV). The HEG1
spectrum is similar to MEG1 but only the Mg XII, Si XIII, and S XV
lines in HEG1 have sufficient  counts to permit line measurements.
Even so, they provide a good cross-check with MEG1 since HEG1
spectral resolution ($\Delta$$\lambda$ = 0.012 \AA~) is superior
to MEG1 ($\Delta$$\lambda$ = 0.023 \AA).

All of the HETG lines are redshifted with exception of S XV (Table 3).
The Ne X line (12 net MEG1 counts) has a redshift of $+$841 ($-$396,$+$593) km s$^{-1}$
and width FWHM $\approx$ 0.04 \AA~ ($\approx$990 km s$^{-1}$).
This width is  nearly twice the MEG1 resolving power. Line broadening
is also present in the {\em XMM-Newton} grating spectra (Sugawara et al. 2008).

Table 3 shows what appears to be a trend toward smaller redshifts in hotter
lines. The individual components of the doublet lines (Ne X, Mg XII, Si XIV)
are not resolved by MEG1 and their redshifts may be slightly overestimated
due to flux contributions from the weaker longer wavelength component.
These doublets appear somewhat broadened in the MEG1 spectrum.
The faint Si XIV line (2.006 keV) is too noisy to permit a
reliable centroid measurement. 
The S XV line (12 net MEG1 counts) is uncontaminated by other lines and quite sharp.
Its measured centroid is $\lambda_{obs}$ = 5.039 ($-$0.007, $+$0.006; 1$\sigma$) \AA~
or $v$ = 0 ($-$417, $+$357) km s$^{-1}$. The uncertainty is within the
limits of MEG1 absolute wavelength calibration accuracy of $\pm$0.011 \AA.
Thus, we measure no significant centroid shift in this hotter S XV line.

\begin{figure}
\figurenum{2}
\includegraphics*[width=9.0cm,angle=-90]{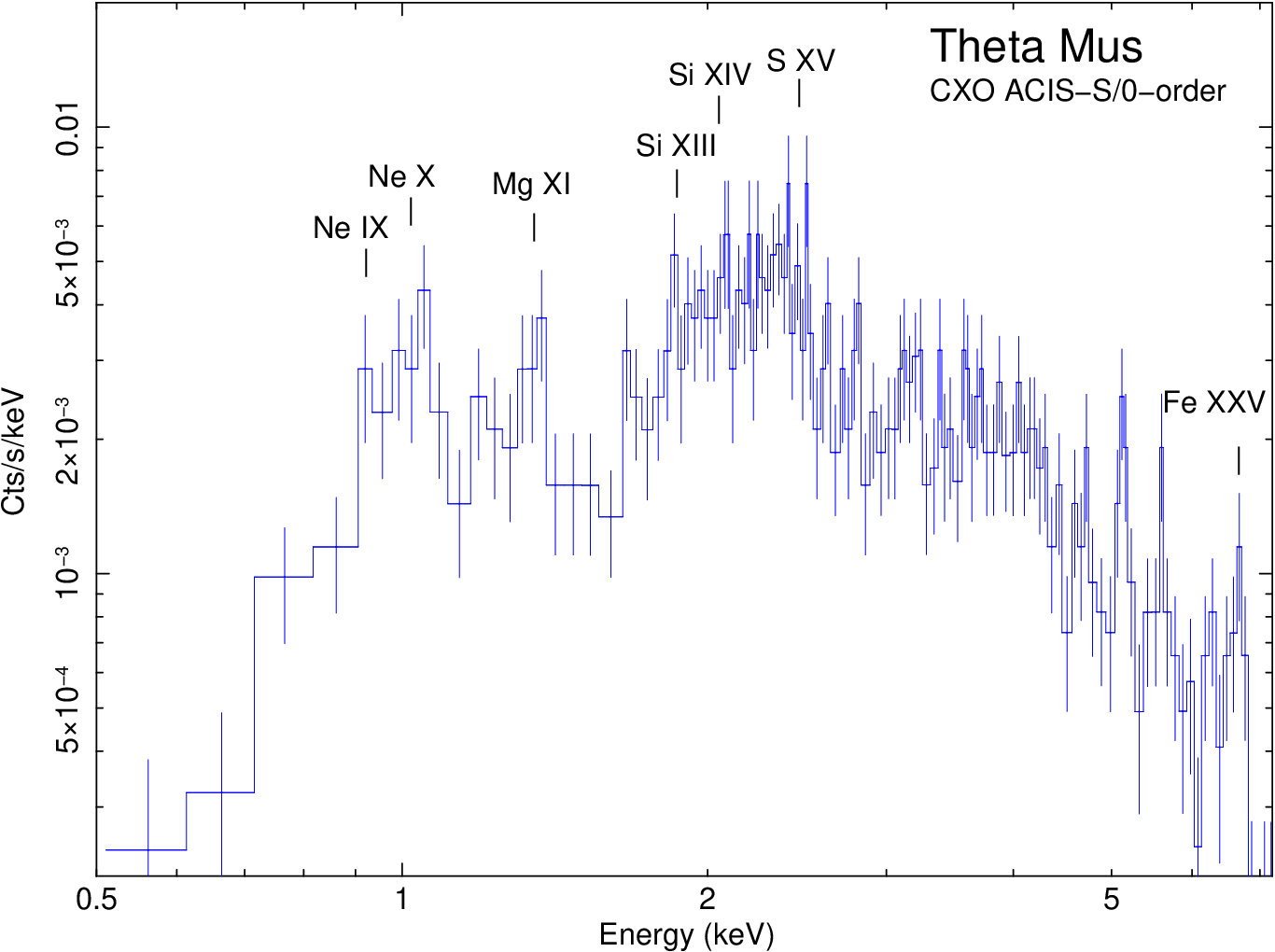} \\
\includegraphics*[width=9.0cm,angle=-90]{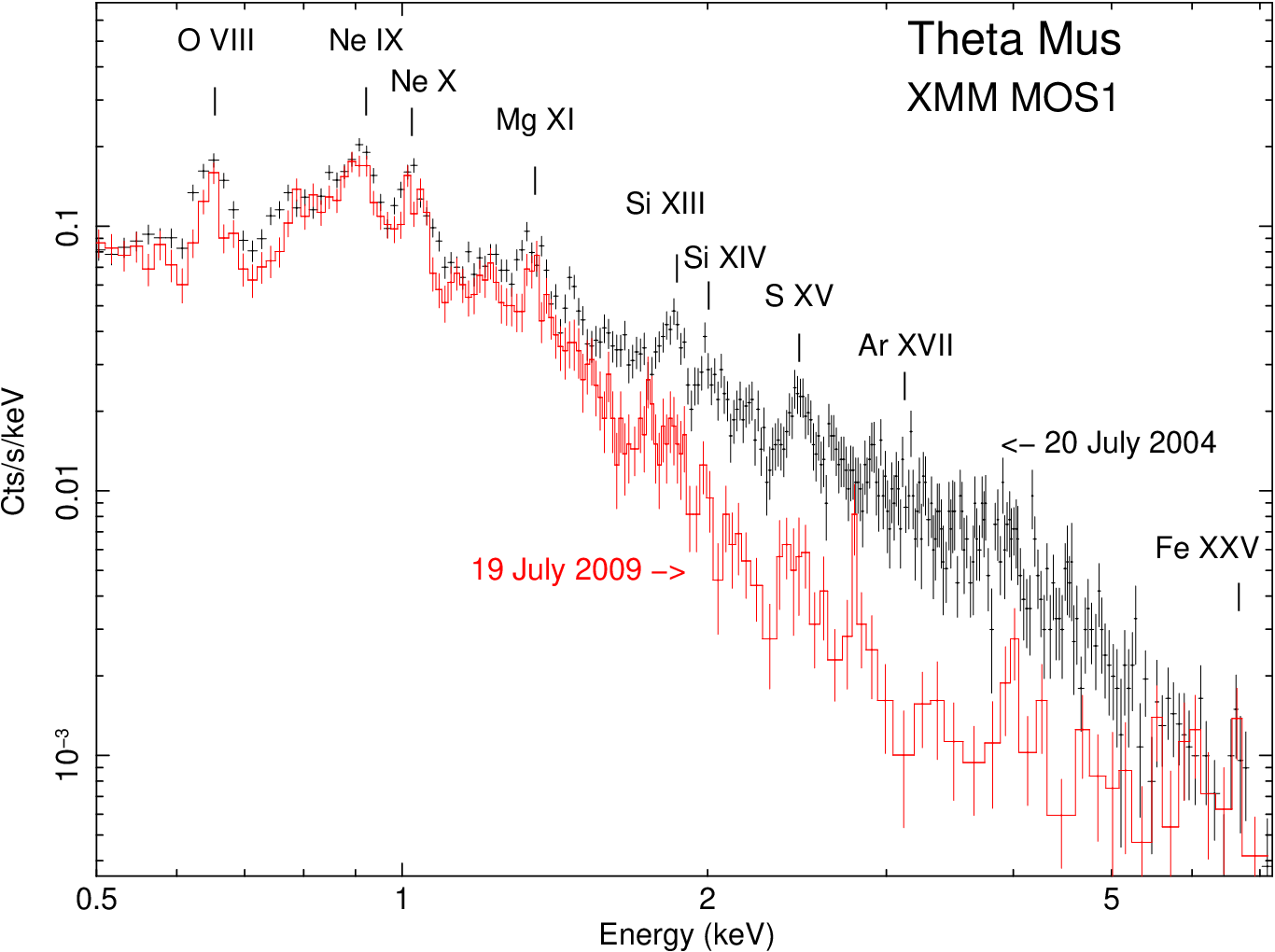}
\caption{Background-subtracted X-ray spectra of $\theta$ Mus (log-log scale). 
{\em Top}:~Summed {\em Chandra} ACIS-S 0-order (1430 cts; minimum
3 cts/bin). The feature at 5.1 keV is unidentified.
{\em Bottom}:~XMM-Newton MOS1 (minimum 10 cts/bin) for
ObsIds 0090030201 (black; 18903 cts) and 0605670201 (red;  6458 cts).
}
\end{figure}

\clearpage

\begin{figure}
\figurenum{3}
\includegraphics*[width=9.0cm,angle=-90]{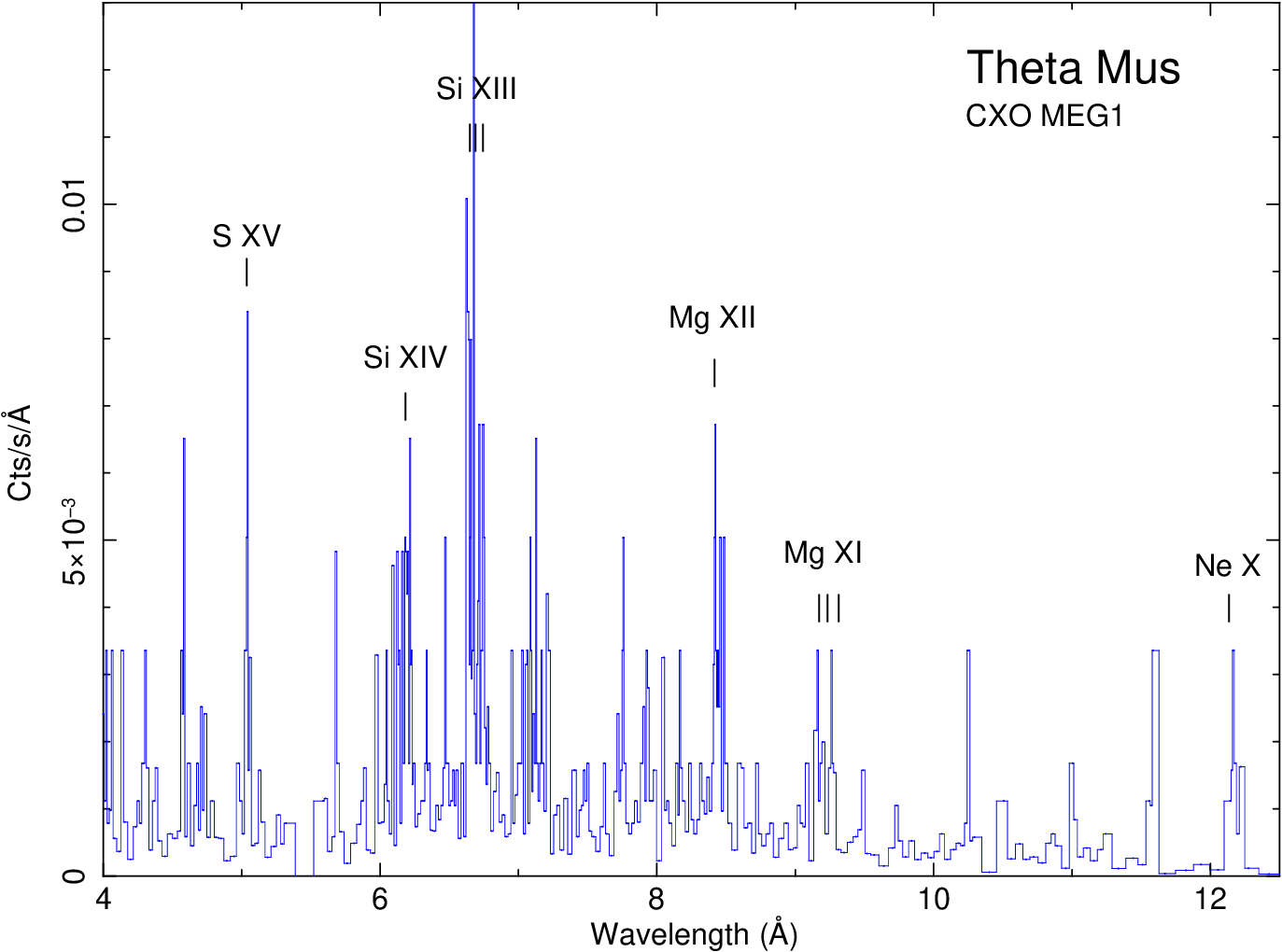} \\
\includegraphics*[width=9.0cm,angle=-90]{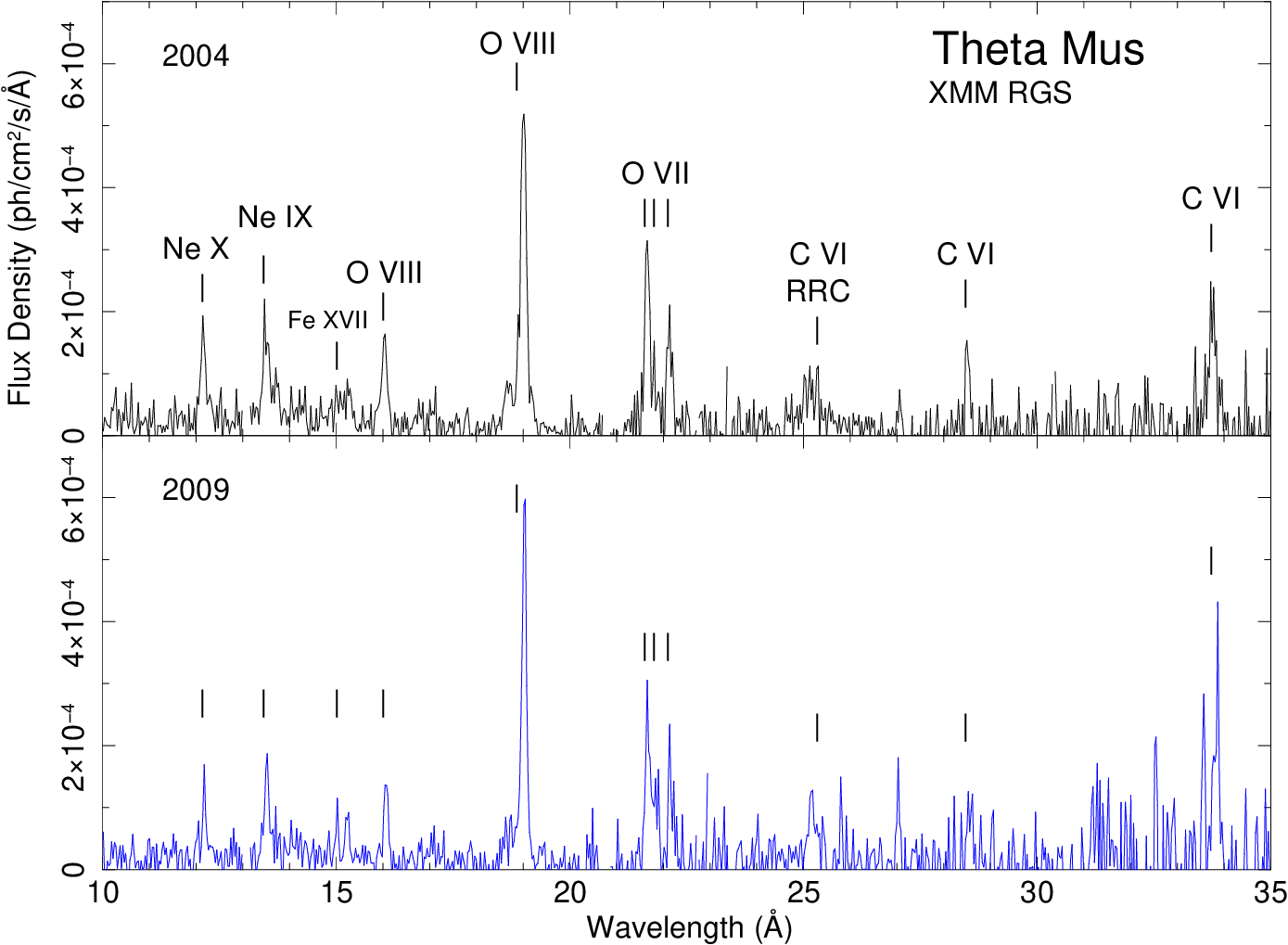}
\caption{$\theta$ Mus 1st order grating spectra. ATOMDB line lab wavelengths are marked.
{\em Top}:~Summed {\em Chandra} MEG1 (1188 cts; lightly rebinned).
Error bars omitted for clarity.
{\em Bottom}: {\em XMM-Newton} RGS1$+$RGS2 (bin width of 0.03 \AA)
for ObsIds 0090030201 (black) and 0605670201 (blue).}
\end{figure}

\clearpage

\begin{deluxetable}{lllll}
\tabletypesize{\scriptsize}
\tablewidth{0pc}
\tablecaption{Spectral Fits of $\theta$ Mus
   \label{tbl-1}}
\tablehead{
\colhead{Parameter}      &   
\colhead{Chandra}        &
\colhead{XMM-Newton}     &
\colhead{XMM-Newton}     &
\colhead{XMM-Newton}
}
\startdata
Instrument                                         & ACIS-S/0-order         & EPIC (pn$+$MOS)    & EPIC (pn$+$MOS)   & EPIC (pn$+$MOS)  \\
ObsId                                              & all                    & 0090030201         & 0090030201        & 0605670201 \\
Year                                               & 2020, 2022             & 2004               & 2004              & 2009      \\ 
Phase ($\phi$)                                     & 0.58 - 0.66            & 0.60 - 0.67        & 0.60 - 0.67       & 0.95 - 0.98 \\
Model\tablenotemark{\scriptsize a}                 & 2T $vpshock$           & 2T $vpshock$       & 2T $vapec$        & 2T $vpshock$     \\
Abundances                                         & varied\tablenotemark{\scriptsize b}   & varied\tablenotemark{\scriptsize c} &  varied\tablenotemark{\scriptsize d} & varied\tablenotemark{\scriptsize e}   \\
N$_{\rm H,1}$ (10$^{22}$ cm$^{-2}$)                & (0.16)                 & 0.16 [0.15 - 0.18] & 0.13 [0.12 - 0.14]  & (0.16)   \\
N$_{\rm H,2}$ (10$^{22}$ cm$^{-2}$)                & ...                    & ...                & ...                 & 17.1 [15.2 - 19.7]  \\     
kT$_{1}$ (keV)                                     & (0.40)                 & 0.51 [0.50 - 0.53] & 0.37 [0.36 - 0.38]  & 0.79 [0.77 - 0.81]       \\
kT$_{2}$ (keV)                                     & 4.67 [4.26 - 5.23]     & 4.34 [4.27 - 4.48] & 4.04 [3.94 - 4.14]  & (4.3)      \\
E$_{edge}$ (keV)                                   & ...\tablenotemark{\scriptsize f} & 0.55 [0.54 - 0.56]  & 0.56 [0.55 - 0.57]          & 0.54 [0.54 - 0.55]        \\
kT$_{edge}$ (eV)                                   & ...\tablenotemark{\scriptsize f} & 4.0  [0.5 - 6.7]    & 3.5  [1.8 - 5.5]            & (4.0)         \\
redshift\tablenotemark{\scriptsize f}              & (0)                      & (2.17e$-$3)                        & (2.17e$-$3)          & (2.17e$-$3)  \\
norm$_{1}$/norm$_{2}$                              & 0.55                  & 1.05                           & 1.57               & 1.86      \\
$\tau_{u,1}$ (s cm$^{-3}$)                         & 2.32e$+$10           & 3.89e$+$11                      & ...                & 4.63e$+$11       \\
$\tau_{u,2}$ (s cm$^{-3}$)                         & 4.67e$+$13           & 15.5e$+$11                      & ...                & ...\tablenotemark{\scriptsize g}              \\
$\chi^2$/dof ($\chi^2_{red}$)                      & 314.8/255 (1.23)     & 2519.7/2281 (1.10)              & 2668.9/2285 (1.17) & 1241.2/1149 (1.08)    \\
F$_{\rm X,abs}$\tablenotemark{\scriptsize h}           & 12.5 [10.6 - 14.0]\tablenotemark{\scriptsize i}   & 16.06 [15.90 - 16.10]           & 16.01 [15.90 - 16.06] & 10.1 [10.0 - 10.2]   \\
F$_{\rm X,2,abs}$\tablenotemark{\scriptsize h}         & 6.10                 & 8.87                            & 8.75                  & 1.06 \\
log L$_{\rm X}$ (ergs s$^{-1}$)                        & 33.31                & 33.36                           & 33.23                 & 33.31 \\
\enddata
\tablecomments{
Based on  XSPEC simultaneous fits of binned background-subtracted spectra. 
The tabulated parameters are
absorption column density (N$_{\rm H}$), plasma temperaure in energy units (kT), 
ratio of cool/hot plasma normalizations (norm$_{1}$/norm$_{2}$),
emission-recombination edge threshold energy (E$_{edge}$) and temperature in energy units (kT$_{edge}$),
upper limit on ionization timescale ($\tau_{u}$),
total absorbed X-ray flux (F$_{\rm X,abs}$), hot component absorbed X-ray flux (F$_{\rm X,2,abs}$), 
and unabsorbed X-ray luminosity L$_{\rm X}$ at d = 2.168 kpc. 
X-ray fluxes and  L$_{\rm X}$ are evaluated in the 0.2-8 keV range.
Square brackets enclose 1$\sigma$ confidence intervals. Values in parentheses 
were held fixed during fitting.
Abundances are fixed at the solar values of Anders \& Grevesse (1989)
except as noted below.}
\tablenotetext{a}{2T $vpshock$ model for {\em Chandra} ACIS-S and {\em XMM-Newton} EPIC 2004 fits 
                  use one absorption component and are of form 
                  $wabs_{1}$$\cdot$($redge$ $+$ $vpshock_{1}$ $+$ $vpshock_{2}$) 
                  where $wabs_{1}$ $\equiv$  N$_{\rm H,1}$. 
                  The EPIC 2009 fit includes a second absorption component
                  $wabs_{2}$ $\equiv$  N$_{\rm H,2}$ that is varied independently and is of form
                  $wabs_{1}$$\cdot$($redge$ $+$ $vpshock_{1}$) $+$ $wabs_{2}$$\cdot$$vpshock_{2}$.
                  2T $vapec$ model is of form $wabs_{1}$$\cdot$($redge$ $+$ $vapec_{1}$ $+$ $vapec_{2}$).}
\tablenotetext{b}{Varied abundances and best-fit values are
                  Fe = 0.27 [0.13 - 0.41] $\times$ solar.}
\tablenotetext{c}{Varied abundances and best-fit values are
                  C = 4.34 [3.27 - 5.71], O = 0.42 [0.40 - 0.48]   ], Ne = 0.74  [0.73 - 0.84],
                  Mg = 0.37 [0.36 - 0.43], Si = 1.14 [1.09 - 1.24], S = 2.18 [2.03 - 2.32] Fe = 0.11 [0.10 - 0.14] $\times$ solar.}
\tablenotetext{d}{Varied abundances and best-fit values are
                  C = (4.34) fixed, O = 0.63 [0.60 - 0.68]   ], Ne = 0.99  [0.93 - 1.07],
                  Mg = 0.46 [0.40 - 0.52], Si = 2.54 [2.38 - 2.70], S = 3.30 [3.01 - 3.60] Fe = 0.10 [0.09 - 0.11] $\times$ solar.}
\tablenotetext{e}{Varied abundances and best-fit values are
                  C = 7.99 [7.58 - 9.33], O = 0.60 [0.54 -  0.67], Ne = 0.50 [0.46 - 0.55],
                  Mg = 0.39 [0.35 - 0.44], Fe = 0.12 [0.11 - 0.14] $\times$ solar.}
\tablenotetext{f}{ACIS-S fits are not sensitive to $redshift$ or $redge$.
                  For EPIC fits redshift was held fixed at [2.17e-3] = 650 km s$^{-1}$  (Sugawara et al. 2008).}
\tablenotetext{g}{Fit did not converge to a stable value.}
\tablenotetext{h}{Flux (0.2-8 keV) is in units of 10$^{-13}$ ergs cm$^{-2}$ s$^{-1}$.}
\tablenotetext{i}{{\em Chandra} ACIS-S 0-order has low effective area below $\approx$1 keV 
                  compared to {\em XMM-Newton} EPIC (Sec. 5.2).}
\end{deluxetable}

\clearpage

\begin{deluxetable}{lcccccc}
\tablewidth{0pt}
\tablecaption{$\theta$ Mus Spectral Lines}
\tablehead{
           \colhead{Ion}                &
           \colhead{E$_{lab}$}          &
           \colhead{$\lambda_{lab}$}    &
           \colhead{$\lambda_{obs}$}    &
           \colhead{$\Delta$$\lambda$}  &
           \colhead{log T$_{max}$}      &
           \colhead{Net Line Flux}  \\
           \colhead{}               &
           \colhead{(keV)}          &
           \colhead{(\AA)}          &
           \colhead{(\AA)}          &
           \colhead{(\AA~[km s$^{-1}$])}          &
           \colhead{(K)}            &
           \colhead{(10$^{-6}$ ph cm$^{-2}$ s$^{-1}$)}             
}
\startdata
\multicolumn{7}{c}{XMM-Newton (RGS)} \\
 O VII$r$ & 0.574  & 21.602                 & 21.663 (2004) & $+$0.061 [$+$847]     & 6.3    & 3.73    \\
 O VII$r$ & 0.574  & 21.602                 & 21.679 (2009) & $+$0.068 [$+$944]     & 6.3    & 3.52    \\
 O VIII   & 0.654  & 18.967/18.973\tablenotemark{\scriptsize a}   & 19.010 (2004) & $+$0.043 [$+$680]     & 6.5  & 7.61    \\
 O VIII   & 0.654  & 18.967/18.973\tablenotemark{\scriptsize a}   & 19.025 (2009) & $+$0.058 [$+$917]     & 6.5  & 6.90     \\
\multicolumn{7}{c}{Chandra (HETG1)} \\
 Ne X     & 1.022  & 12.132/12.137\tablenotemark{\scriptsize a} & 12.166 (MEG) & $+$0.034~ [$+$841] & 6.8    & 9.95  \\
 Mg XII  & 1.473  &  8.419/8.425\tablenotemark{\scriptsize a}   &  8.437 (HEG) & $+$0.018~ [$+$641] & 7.0    & 1.87  \\
Si XIIIr & 1.865  &  6.648 &  6.655 (MEG) & $+$0.007~ [$+$316] & 7.0    & 2.09  \\
Si XIIIf & 1.840  &  6.740 &  6.746 (MEG) & $+$0.006~ [$+$267] & 7.0    & 1.51  \\
 S XV    & 2.460  &   5.039 & 5.039 (MEG,HEG) & 0~~~~~~~~~~[0]     & 7.2    & 3.70  
\enddata
\noindent Notes: 
Laboratory wavelengths ($\lambda_{lab}$) and maximum line power temperatures (T$_{max}$) 
are from the ATOMDB database (www.atomdb.org). Line measurements are based on the
summed MEG1 spectra from all five observations except for Si XIII$r$ (resonance) and 
Si XIII$f$ (forbidden) lines which were only measurable in ObsIds 24497 and 24498.
Net photon (ph) line fluxes are continuum-subtracted. Typical flux uncertainties are 20\%.
\tablenotetext{a}{Blended doublet.}
\end{deluxetable}

\clearpage

\subsection{XMM-Newton}

We compare here the {\em XMM-Newton} spectra with {\em Chandra}. 
The comparison is useful because
{\em XMM-Newton}  provides better low-energy sensitivity 
and spectra were obtained at nearly the same phase as
{\em Chandra} and at opposite phase where {\em Chandra} data are lacking.

{\em Undispersed EPIC Spectra}.~
Figure 2-bottom shows EPIC MOS1 spectra for both 
{\em XMM-Newton} observations. These spectra give reliable broad-band fluxes and 
estimates of basic spectral parameters (N$_{\rm H}$, kT).
Table 2 compares fits of the EPIC  and {\em Chandra} ACIS-S 0-order spectra. 
The 2T $vpshock$ fits of the ACIS-S and 2004 EPIC 
spectrum obtained at about the same phase give similar results 
but the ACIS-S fits place less emission measure in the cooler 
plasma component. This is largely due to the lower ACIS-S
sensitivity at energies below 1 keV. 
This difference is apparent in Figure 2 which reveals
significant emission below $\approx$1 keV in the MOS
spectra such as the  O VIII line which is not detected by 
{\em Chandra}. Thus, ACIS-S fits yield a lower measured
flux than EPIC at $\phi$ $\approx$ 0.65 (Table 2). 
EPIC spectral fits are slightly improved  
in the range 0.49 - 0.55 keV near the C VI 
radiative recombination continuum (RRC) feature
by including a recombination edge ($redge$) component 
as did Sugawara et al. (2008).
The RRC feature is faint in the EPIC spectra
but clearly visible in RGS. 
The $redge$ component does not affect ACIS-S fits 
since ACIS-S detects very little flux near 0.5 keV.

Table 2 includes a fit of the 2004 EPIC spectra using a
2T  $vapec$ optically thin plasma model for comparison
with 2T $vpshock$. The two models give similar results
but 2T $vpshock$  yields 
a slightly higher absorption N$_{\rm H}$ and thus a 
higher L$_{x}$. Also the 2T $vpshock$ model converges to a more 
stable value of the  C abundance which is above solar and consistent
with that obtained from 2T $vpshock$ fits 
by Sugawara et al. (2008). 

The 2009 {\em XMM-Newton} observation at  
phase $\phi$ =  0.95 - 0.98  provides crucial variability 
information. At lower energies the 2009 EPIC spectrum is 
quite similar to the 2004 EPIC spectrum but a modest flux change 
of $\leq$20\% below 1.5 keV is not ruled out.
More obvious is the decrease in emergent flux
at  higher energies $E$ $\approx$2 - 5 keV 
in the 2009 EPIC spectrum,  but little if any change
at the Fe XXV energy (6.67 keV).
Fits of the 2009 EPIC spectra give an absorbed broad-band flux 
F$_{x,abs}$(0.2-8 keV) = 9.85 [9.41 - 10.1] $\times$ 10$^{-13}$ ergs cm$^{-2}$ s$^{-1}$. 
This is only 61\% of the 2004 value. The hard-band (2-8 keV) flux
dropped by more than a factor of 3 compared to the 2004 observation
and accounts for most of the change.
The emission measure and observed flux in the 2009 observation are
dominated by the cooler plasma component.
Even so, hot plasma traced by Fe XXV (T$_{max}$ $\approx$ 63 MK)
is also present but the 2009 EPIC spectra do not
tightly constrain the hot-component temperature.

The decrease in hard-band flux could  be due to a change in
the intrinsic X-ray spectrum or an increase in the absorption of the hotter 
plasma component. Although some intrinsic spectral change is not excluded
an increase in absorption due to a change in viewing geometry
over orbital phase is more easily reconciled with the CW picture. 
In that interpretation higher absorption is expected when the WR star passes
in front ($\phi$ $\approx$ 0) and the line-of-sight 
toward the hottest CW shock plasma near the line-of-centers passes through 
the dense metal-enriched WR wind. Lower absorption is anticipated when the 
OV star passes in front ($\phi$ $\approx$ 0.5) and the hotter CW shock plasma
is viewed through a low-density cavity formed by the OV star wind.

To test this hypothesis a separate absorption component (N$_{\rm H,2}$)
was applied to the hotter plasma component in the 2T $vpshock$ model 
of the 2009 EPIC spectra. This model provides a good fit (Table 2) 
and requires high but loosely constrained absorption 
N$_{\rm H,2}$ $\sim$ 10$^{23}$ cm$^{-2}$.
The unfolded spectrum (Fig. 4) shows that the hotter component dominates 
the observed spectrum above $\approx$3 keV and accounts for the Fe XXV emission. 
But the crossover energy depends on the plasma temperatures and absorption 
column density. Factors affecting the absorption are discussed below (Sec. 6.4.1).

This absorbed hot component  is strikingly similar to that 
derived from {\em XMM-Newton} observations of $\gamma^2$ Vel 
at low-state (Fig. 5 of Schild et al. 2004). Given the 
similarity it is likely that the hotter plasma originates 
in the CW interaction region near the line of centers
where high temperatures and densities are expected.
The key question is whether the CW region is in the SB
system or the  SB$+$OI system (Sec. 6.4.1).

\begin{figure}
\figurenum{4}
\includegraphics*[width=9.0cm,angle=-90]{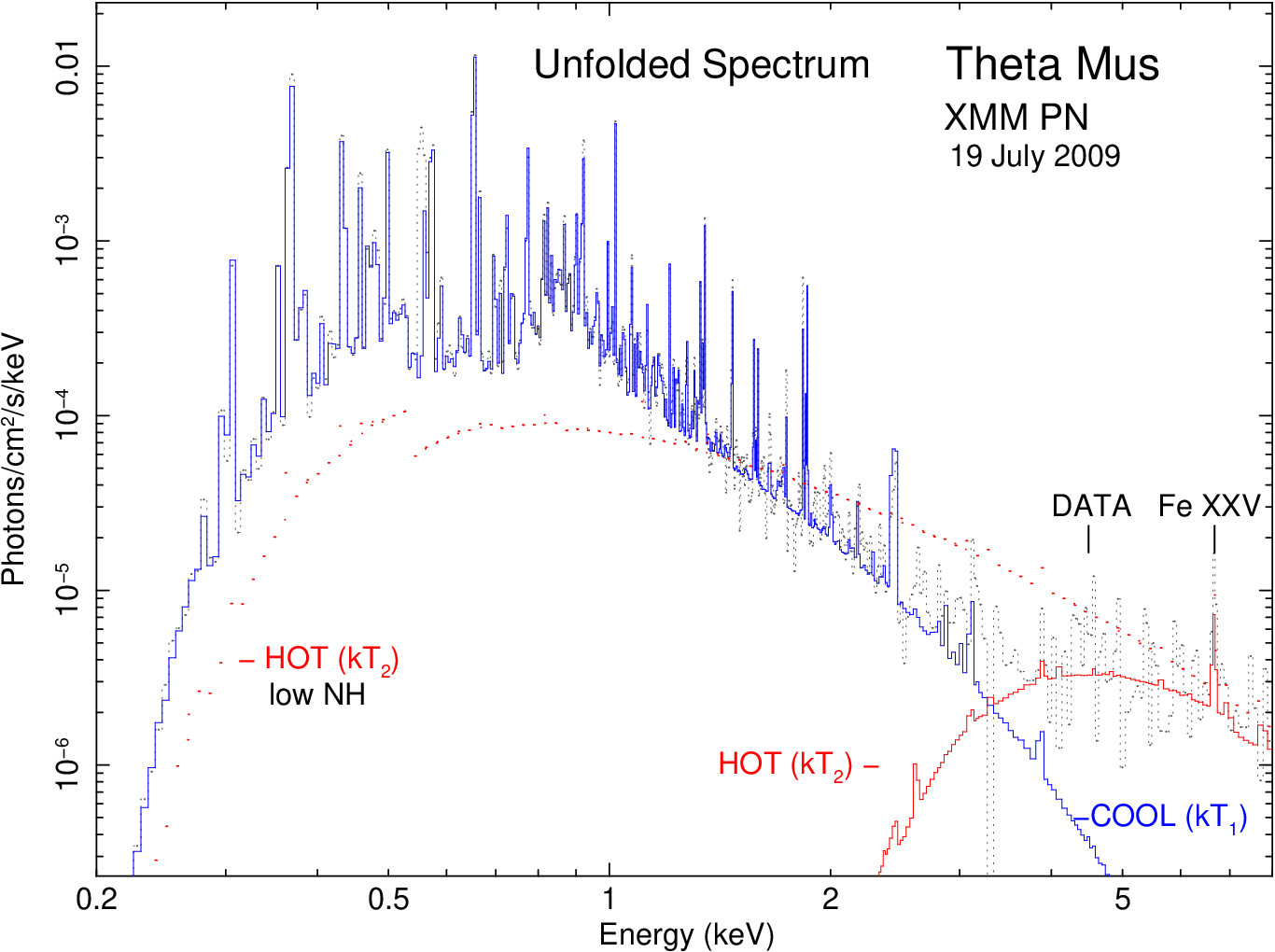}  \\
\caption{Unfolded {\em XMM-Newton} ObsId 0605670201 EPIC pn spectrum of 
$\theta$ Mus (gray dotted) showing cool and hot 2T $vpshock$ model components (Table 2).
The dotted red line shows the heavily absorbed hot component as it
would appear if viewed under the same low absorption as the cool component.
}
\end{figure}

{\em RGS Grating Spectra}.~
Although detailed analysis of the {\em XMM-Newton} RGS grating spectra
is beyond the scope of this work, we briefly summarize key results.
The RGS spectra (Fig. 3) are remarkably similar even though they were
obtained five years apart at opposite phases. Low-energy lines down to
C VI at $\lambda_{lab}$ = 33.734 \AA~  were detected. 
The C VI RRC feature at 25.30 \AA~  is present in both spectra.
Our analysis reveals that
brighter lines are clearly broadened and redshifted in {\em both} observations 
with slightly higher shifts measured in the 2009 observation (Table 3). The strong 
O VIII line has width FWHM $\approx$ 1900$\pm$400 km s$^{-1}$ and is redshifted
to $\approx$$+$700 - $+$900 km s$^{-1}$.
The O VIII line fluxes of the two observations differ by no more than
$\approx$15\% suggesting that O VIII  emission is nearly stable over 5 years.

The RGS spectra detect the O VII and Ne IX He-like triplets  
consisting of the closely-spaced resonance ($r$), intercombination ($i$), 
and forbidden ($f$) lines. In hot stars with strong UV radiation fields the 
line flux ratio $R$ = $f/i$ is not a reliable electron 
density diagnostic but can constrain the line formation distance from the star
and $G$ = ($f + i$)/$r$ is temperature sensitive (Gabriel \& Jordan 1969; 
Blumenthal, Drake, \& Tucker 1972; Pradhan 1982; Porquet et al. 2001). 
The triplet components are well separated 
in O VII but not Ne IX. For the longer 2004 observation with highest signal-to-noise ratio
we obtain O VII flux ratios (normalizing the $r$ line flux to unity)
$r$:$i$:$f$ = 1.00:0.17:0.50 giving 
$R$ = 3.0 [2.0 - 4.9] and $G$ = 0.7 [0.5 - 0.9].
The large uncertainty range for $R$ is due to the uncertainty
in measuring the flux of the faint $i$ line.
In the case of low collisional excitation where the 
electron density is much less than the critical density
($n_{e}$ $\ll$ $n_{c}$) and stellar photoexcitation is
well below the critical photoexcitation rate 
($\phi_{*}$ $\ll$ $\phi_{c}$)
the limiting values for O VII are
$R_{o}$ = 3.95 and $G_{o}$ = 1.0 (Pradhan 1982).
Taking the uncertainty range of $G$ into account a
value $G$ $<$ $G_{o}$ is not ruled out.
If $G$ $<$ $G_{o}$ the plasma
is not in ionization equilibrium at the
line formation temperature of O VII (Pradhan 1982).
Sugawara et al. (2008) have argued for NEI conditions.

A lower limit on the O VII formation distance  r$_{OV}$
from the O6-7V star can be obtained using the 
lower bound on $R$ as was done for 
$\gamma^2$ Vel (Skinner et al. 2001; Schild et al. 2004).
The approach assumes the electron density is
subcritical ($n_{e}$ $\ll$ $n_{c}$) and that
photoexcitation is responsible for decreasing
$R$ to a value below $R_{o}$.
Using the lower bound $R$ $\geq$ 2.0 for O VII gives a 
$R$/$R_{o}$ $\geq$ 0.5. Assuming T$_{eff}$ $\approx$ 38,000 K
for the O6-7V star (Weidner \& Vink 2010) and scaling the 
value of $\phi_{*}$/$\phi_{c}$ in Table 2 of Blumenthal et al. (1972) 
for T$_{eff}$ = 10$^{5}$  down to 38 kK gives a lower
limit r$_{OV}$ $\gtsimeq$ 70 $R_{OV}$. This bound is
a first approximation and a more refined estimate
would require detailed calculation of the stellar
radiation fields. But it is clear that if photoexcitation 
dominates over collisions then the O VII triplet does
not form close to the O6-7V star.

\subsection{Summary of Spectral Analysis}
The X-ray emission of $\theta$ Mus consists of a weakly-absorbed nearly steady cool
component at kT$_{1}$ $\approx$ 0.4 - 0.7 keV 
plus a hotter component at kT$_{2}$ $\gtsimeq$ 4 keV whose observed 
flux is variable on a timescale of $\leq$5 years. The hotter plasma 
contributed a greater fraction of the total X-ray flux at orbital phase 
$\phi$ $\approx$ 0.65 than  at $\phi$ $\approx$ 0.96. 
Emission lines are broadened to FWHM $\geq$ 1000 km s$^{-1}$ due to 
bulk plasma motion. Low-temperature emission lines are redshifted to 
velocities as high as $v$ $\approx$ $+$800 - $+$900 km s$^{-1}$.
But no significant centroid shift was measured in {\em Chandra} HETG
spectra of the S XV line which traces warmer plasma 
that would be located nearer to the line-of-centers in CW shock binaries. 
The strong O VII forbidden line in the 2004 {\em XMM-Newton}
RGS spectrum can be explained if O VII forms in cool low-density 
plasma located far from the O6-7V star. The significant decline in
hot component flux can be attributed to increased  
absorption of the hot component at phase $\phi$ $\approx$ 0.95 - 0.98
(WR in front). The $\theta$ Mus X-ray luminosity for the  
observations analyzed herein is
log L$_{x}$(0.2 - 8 keV) = 33.22 - 33.36 ergs s$^{-1}$ (Table 2),
close to the {\em Einstein} value log L$_{x}$ = 33.2 ergs s$^{-1}$.
It is also similar to other massive WC binaries (Pollock 1987) including
$\gamma^2$ Velorum (log L$_{x}$ = 33.06 - 33.17; Schild et al. 2004). 

\section{Discussion}

\subsection{Comparison with $\gamma^2$ Velorum}

The X-ray observations of $\theta$ Mus show some striking
similarities with the 78-day WC8$+$O7.5 binary 
$\gamma^2$ Velorum  studied by Willis, Schild, \& Stevens (1995), 
Stevens et al. (1996), Skinner et al. (2001), and Schild et al. (2004).
In addition to their similar spectral types both systems have a 
cool plasma component that is viewed under low absorption
and appears to be nearly steady with orbital phase.
The C VI RRC feature in both systems points to an extended region
where carbon ions that likely originate in the WC star wind recombine 
with free electrons. Extended cool plasma located far out in the system
is also implicated by strong forbidden lines in He-like triplets
for both stars. In addition,
hotter plasma including the Fe XXV line evidently viewed 
through high absorption at certain phases is present, as
expected for CW shock emission near the line-of-centers. 
Line broadening is detected in both sytems with 
widths FWHM $\geq$ 1000 km s$^{-1}$.
The X-ray luminosity of  $\gamma^2$ Vel is similar to  $\theta$ Mus.
However, one notable difference is that the {\em low}-temperature
X-ray emission lines of $\theta$ Mus are redshifted but no significant line 
shifts have been detected in $\gamma^2$ Vel (Skinner et al. 2001). 
The unusual line shifts (or lack thereof) are not easily reconciled 
with colliding wind theory.

\subsection{X-ray Emission from the O Stars}
Although intrinsic X-rays from the WC6 star are not anticipated (Sec. 1)
the X-ray emission of $\theta$ Mus could include
intrinsic emission from the OV or OI stars.
As an example, soft emission from the O9III star 
$\theta$ Mus B lying south of $\theta$ Mus is faintly detected 
by {\em Chandra} (Fig. 1). X-ray grating observations of 
$\zeta$ Ori (O9.7Ib) show cool plasma that is consistent
with radiative wind shocks and heavily suppressed forbidden
lines in O VII and Ne IX (Waldron \& Cassinelli 2000; Raassen et al. 2008).
The suppressed $f$ lines imply cool plasma formation close to the
star, clearly different from $\theta$ Mus. The $\zeta$ Ori
X-ray luminosity is log L$_{x}$ = 32.14 ergs s$^{-1}$ (Raassen et al. 2008).
If L$_{x}$ of the O9.5I star in the $\theta$ Mus system
is similar then it would account for $<$10\% of the observed value.

It is also unlikely that intrinsic O6-7V star emission alone  
can account for the observed L$_{x}$, the hotter plasma traced 
by the Fe XXV line, and the redshifted lines.
Main sequence O-type stars are generally soft X-ray sources
and a recent study based on {\em XMM-Newton} data and {\em Gaia} 
distances yields the correlation 
log (L$_{x}$/L$_{bol}$) = $-$6.6$\pm$0.4 (G\'{o}mez-Mor\'{a}n \& Oskinova 2018). 
Adopting a typical O6-7V value
log (L$_{bol}$/L$_{\odot}$) =  5.65 gives 
L$_{x}$ = 0.11 (0.045 - 0.28) L$_{\odot}$ or 
log L$_{x}$ = 32.6 (32.23 - 33.03) ergs s$^{-1}$ which  
is less than the observed value log L$_{x}$ = 33.3 ergs s$^{-1}$. 
Thus, intrinsic emission from the O stars could account
for some, but not all, of the observed X-ray luminosity.
But if the O-star contribution were due to radiative 
wind shocks then one would expect low-temperature lines to be 
blueshifted (not redshifted) and the presence of hot plasma in $\theta$ Mus
is at odds with radiative wind shock predictions.

\subsection{Colliding Wind Predictions}

The presence of redshifted {\em low-temperature} emission lines 
in $\theta$ Mus  at phase $\phi$ $\approx$ 0.6 when the OV star 
should  be nearly in front of the WC star is unexpected if the cooler 
X-ray plasma arises between the WC and O star in the SB system (Sec. 2).
As suggested by Sugawara et al. (2008) the redshifted lines 
could be reconciled with the CW interpretation if
the CW shock lies between the SB system and the OI star
lying further away in a bound orbit. But regardless of
whether a CW shock is located in the wide SB$+$OI system
or the SB system it is evident from the low absorption of
cooler plasma, C VI RRC feature, and O VII $f$/$i$ line
ratio that some cool plasma is located far out in an
extended region where the wind density is low. 
On the other hand
the drop in flux of the hotter plasma component on a
timescale of $\leq$5 years is compatible with  
the 19-day SB system but would be rather fortuitous if the 
SB$+$OI system period is significantly greater than 40 years.
Below we compare CW predictions for SB system
and the wide SB$+$OI system.

{\em SB System}.~
The location of the stagnation point on the line-of-centers between
the WC6 and O6-7V stars is determined by the wind momentum ratios
(eq. [1] of Stevens et al. 1992)~

\begin{equation}
\left[\frac{\dot{M}_{WC} V_{WC}(r_{1})}{\dot{M}_{OV} V_{OV}(r_{2})}\right]^{1/2}  = \frac{r_{1}}{r_{2}} \equiv \mathcal{R}
\end{equation}
where $r_{1}$ and $r_{2}$ are the respective distances from the 
center of the WC and O stars to the stagnation point and
$r_{1}$ $+$ $r_{2}$ = $D$ is the binary separation.
Mass loss parameters for WR and O stars are rather uncertain
but for purposes of an estimate we adopt the 
WC6 mass loss rate log $\dot{M}_{WC}$ = $-$4.74 M$_{\odot}$ yr$^{-1}$ 
and terminal wind speed $V_{\infty,WC}$ = 2060 km s$^{-1}$ of
Nugis \& Lamers (2000).  For the O6-7V star we assume
log $\dot{M}_{OV}$ = $-$6.3 M$_{\odot}$ yr$^{-1}$
and  $V_{\infty,OV}$ = 2450 km s$^{-1}$ (Lamers \& Leitherer 1993).
These adopted values give $\mathcal{R}$ = 5.5 and place the stagnation point at the contact
discontinuity close to the O6-7V star at 
$r_{2}$ = 0.15$D$. Using assumed masses M$_{WC6}$ = 12 M$_{\odot}$
(Hill et al. 2002) and M$_{OV}$ = 30 M$_{\odot}$ (Weidner \& Vink 2010),
the computed semi-major axis is $a$ = 0.49 AU = 105 R$_{\odot}$
for P$_{orb}$ = 19.1325 d. A slightly lower mass for the OV star is
possible (Moffat \& Seggewiss 1977) but would not significantly alter
the computed value of the semi-major axis. Typical radii for WC6 stars
are R$_{WC6}$ $\approx$ 3 R$_{\odot}$ (Sander et al. 2012) and
for O7V stars R$_{OV}$ $\approx$ 10 R$_{\odot}$ (Lamers \& Leitherer 1993).
This gives $r_{2}$ $\approx$ 32 R$_{\odot}$ $\approx$ 3 R$_{O7V}$.
The above results assume the winds of both stars have reached terminal
speed before colliding. This would be the case for the WC6 star assuming
a typical hot star wind velocity profile
$V(r)$ = $V_{\infty}[1 - (R_{*}/r)]^{\beta}$ 
with  $\beta$ $\approx$ $+$0.7 - $+$1.0 (Krti\v{c}ka \& Kub\'{a}t 2011). 
But the O star wind would only be at about 70\% of its terminal speed
which would tend to push the stagnation point even closer to
(or perhaps onto) the O star surface. Thus the hottest plasma would
be located near the line-of-centers close to the OV star.

Theory provides analytic expressions for the CW shock X-ray
luminosity (L$_{x,cw}$) that depend on mass loss parameters,
binary separation, stellar radii, and whether the shock is
adiabatic or radiative (e.g. Usov 1992; Stevens et al. 1992).
But previous studies of WR$+$OB systems have not always yielded 
good agreement between the observed L$_{x}$ and predictions
(e.g. Skinner et al. 2019b). The discrepancies may be in part
due to inaccurate mass loss rates resulting from
inhomogeneous (i.e. clumped) winds (Cherepashchuk 1990; Zhekov 2012).
Also for wide WR binaries where the winds have reached terminal speeds
before colliding, theory predicts that L$_{x,cw}$ should scale inversely with
orbital separation $D$. But no such change was detected in
the hard-band X-ray emission of $\gamma^2$ Vel at two different orbital
phases where $D$ is thought to have changed (Schild et al. 2004).  It
thus seems that not all WR binaries faithfully adhere to the theoretical
L$_{x,cw}$ $\propto$ 1/$D$ scaling relation.

For wide binaries with periods more than a few days the CW
shock is expected to be adiabatic but for short-period
systems it can be radiative. In the latter case most of
the shock energy is radiated away (tending to increase
L$_{x,cw}$) and the shocked plasma is nearly isothermal.  
A rough gauge on whether the shock is adiabatic or
radiative is given by the ratio of the cooling time to
escape time $\eta$ = $t_{cool}$/$t_{esc}$ where
$\eta$ $\ll$ 1 is typical for radiative shocks and 
$\eta$ $\gtsimeq$ 1 for adiabatic shocks. 
The wide SB$+$OI system is clearly in the adiabatic regime
but the close SB system could be either adiabatic or radiative. 
For the $\theta$ Mus SB system we obtain
$\eta$ $\approx$ 1.3 (eq. [8] of Stevens et al. 1992)
which favors the adiabatic regime, but marginally so.
However, if the WR wind shocks onto the surface of
the O star companion then radiative
cooling can be important even if $\eta$ $>$ 1 (Stevens et al. 1992).

If we assume,  as suggested above, that the WC star wind impacts the surface of the
O6-7V star (with the shock cone wrapping around it) then the CW luminosity predicted 
for an adiabatic shock (eq. [81] of Usov 1992)
is more than an order of magnitude less than the observed L$_{x}$. But if
the shock is radiative then (eq. [80] of Usov 1992)
\begin{equation}
L_{x,cw} =  \frac{1}{8}\left[\frac{R_{\rm OV}}{D}\right]^2L_{wind,WC}
\end{equation}
where the wind luminosity is
$L_{wind,WC}$ = (1/2)$\dot{M}_{WC}V_{\infty,WC}^2$ and we have 
assumed spherical homogenous winds with mass density
$\rho_{\infty}$ =  $\dot{M}_ {WC}$/(4$\pi$$D^{2}$$V_{\infty,WC}$).
Using the same mass loss, stellar, and orbital parameters as above 
gives log $L_{x,cw}$ = 33.8 ergs s$^{-1}$, or about $+$0.5 dex 
higher than observed. This difference could  potentially be explained
if the adopted WC mass loss rate is too high as may be the case if
the wind is clumped. 
Thus if the WR wind shocks onto the O6-7V star
the predicted $L_{x,cw}$ for a radiative shock is in better agreement 
with the observed value. Even so, the above $L_{x,cw}$ estimates
are only approximate since the Usov (1992) formulation assumes
a bremsstrahlung X-ray spectrum and does not account for line emission.

{\em SB$+$OI System.}~
A similar analysis to above can be carried out for the SB$+$OI system.
Because of the much wider separation the CW shock would be adiabatic
and the winds would be at terminal speeds before colliding. 
In Eq. (1) we assume that the 
combined SB wind is dominated by the powerful WC6 wind and use its wind
speed and mass loss rate.  For the OI wind we adopt
$\dot{M}_{OI}$ = 3.2 $\times$ 10$^{-6}$ M$_{\odot}$ yr$^{-1}$ 
and $V_{\infty,OI}$ = 1800 km s$^{-1}$ (Lamers \& Leitherer 1993).
For these wind parameters $\mathcal{R}$ = 2.55 and the contact 
discontinuity lies closest to the OI star at an offset of 0.28$D$. 
To estimate the 
deprojected separation $D$ we assume an arbitrary angle of 
45$^{\circ}$ between the line-of-sight and the major axis of 
the SB$+$OI system which gives  $D$ = 141 AU. In that case
the discontinuity is offset from the OI star along the 
line-of-centers by 40 AU.
The predicted X-ray luminosity of the CW shock between the 
SB system and OI  star is (summing results of eqs. [89, 95] of
Usov 1992) log $L_{x,cw}$ $\approx$ 33.6 ergs s$^{-1}$ 
which is within a factor of two of the observed value for
$\theta$ Mus. This is rather good agreement and an adiabatic
CW shock in the SB$+$OI system could account for the 
observed L$_{x}$ unless $D$ is much greater than assumed above.

\subsection{Questions and Challenges for the Colliding Wind Interpretation }

\subsubsection{Where is the Hot Plasma Located?}

If the OI supergiant is in a long-period orbit around the SB system then
two CW shock regions may be present: one in the SB system and another
in the SB$+$OI system. If so their contributions would be superimposed 
in the X-ray spectra. This creates some ambiguity as to where the 
X-ray emission originates.

As we have shown, the cooler plasma is located in an extended
region surrounding the $\theta$ Mus system. The 
redshifted lines in the cooler plasma component that show little
change at opposite SB orbital phases can be explained
if the cooler plasma originates in a CW shock in the wide
SB$+$OI system with the OI star being further away (Sugawara et al. 2008).
Even if the above picture is correct, it does not guarantee that
the hottest plasma also  originates in the SB$+$OI system.
The key diagnostic of the hot plasma is
the Fe XXV line at 6.67 keV. Unfortunately this line is
not bright enough to obtain a reliable centroid measurement
in the {\em Chandra} grating spectra and is not captured
by {\em XMM-Newton} RGS. So we do not know if it is 
persistently redshifted, as are the lines tracing cooler plasma. 

There are two clues pointing to the SB system as the
source of the hotter plasma. One is the 
hard-band variability timescale of $\leq$5 years.
This timescale is not problematic for the 19-day SB
orbit but may be for the wide SB$+$OI system whose orbital
period is not known but is certainly $\gtsimeq$40 years.

The second clue is the high absorption toward the hot component 
N$_{\rm H,2}$ $\sim$ 10$^{23}$ cm$^{-2}$ inferred from
fits of the 2009 EPIC spectrum. This value is similar to
that determined for $\gamma^2$ Vel 
and was attributed to a CW shock between the WC and OIII
stars (Schild et al. 2004). 
CW theory predicts a characteristic absorption column density near
orbital quadrature (Stevens et al. 1992)
\begin{equation}
N_{\rm H,cw} \sim 5\times10^{22} \left[\frac{\dot{M}_{-6}}{V_{wind,1000}} \right]\left[\frac{1 + \mathcal{R}}{\mathcal{R}}\right]D^{-1}_{12}~{\rm cm}^{-2}
\end{equation}
where $\dot{M}_{-6}$ (units of 10$^{-6}$ M$_{\odot}$ yr$^{-1}$)
and V$_{wind,1000}$ (units of 1000 km s$^{-1}$) correspond to
the dominant wind, $\mathcal{R}$ is the wind momentum ratio of the
dominant to weaker wind (eq. [1]) and $D_{12}$ is the separation
in units of 10$^{12}$ cm. For the SB system $\mathcal{R}$ $\approx$ 5.5
and for the SB$+$OI system $\mathcal{R}$ $\approx$ 2.55, as computed above.
The key discriminant is the separation $D$ which is at
least 100 times larger in the  SB$+$OI system.
That leads to a much smaller absorption estimate
N$_{\rm H,cw}$ $\sim$ 10$^{19.8}$ cm$^{-2}$ compared
to the SB system $\sim$ 10$^{22.6}$ cm$^{-2}$.
The wind absorption is abundance dependent and will vary over 
the orbit as the line-of-sight through the wind interaction
region changes. But the much larger characteristic value of N$_{\rm H,cw}$ 
predicted for the SB system is clearly in better 
agreement with that needed to explain the decrease in
hard-band flux in the 2009 {\em XMM-Newton} observation.

\subsubsection{The Mystery of the Missing Blueshifted Lines}

Even if the persistent redshifted low-temperature lines originate in a 
CW shock in the wide SB$+$OI system, that does not explain the absence 
of blueshifted lines near SB orbital phase $\phi$ $\approx$ 0.5-0.6 
(OV star in front) where they should be seen if detectable CW
emission emerges from the  SB system.
This is not the first time questions about undetected
line shifts have arisen in massive WR binaries (Sec. 6.1).
A few possible explanations for the no-show blueshifts
are considered below, all being somewhat speculative.

If there were no {\em detectable} CW shock emission from
the SB system then the observed X-ray emission
would be entirely attributed to a CW shock in the
long-period SB$+$OI system. If the OI star is in a
bound orbit and $P_{orb,OI}$ $\gg$ 40 years then 
its orbital motion over the time interval of 18 years
spanned by the X-ray observations might not be sufficient
to cause line centroids to shift from red to blue. In fact no
line shifts from red to blue were detected 
(Table 3) so a period $\gg$40 years is 
consistent with the X-ray results. 
Obviously, if blueshifted lines are 
not expected then their absence is no longer a mystery.
But in this picture the SB system would {\em not} contribute
significantly to the X-ray emission and this would be
quite exceptional if the WR companion is indeed a OV star.
Other WC$+$O binaries are bright X-ray sources
well above the detection limits of the observations
discussed here (Pollock 1987). Also if the SB system
were X-ray faint (or quiet) then the hot heavily-absorbed plasma 
would be attributed to the wide SB$+$OI system. But as shown  
above the high absorption needed to account for the decrease
in hard-band flux is much greater than predicted by CW theory
for the wide SB$+$OI system. We thus conclude that the
hypothesis of no detectable CW X-rays from the SB system
could explain the lack of blueshifted lines but it does
not square well with what is known about X-ray emission of
other WC$+$O binaries or CW theory absorption predictions.

Line centroid shifts originating in high-velocity shocked gas
in the SB system would be difficult to detect if
the SB orbit is viewed at low inclination (face-on)
or if the shock cone opening angle is large and the flow
velocity along the line-of-sight is small. But on the 
contrary optical radial velocity variations imply
a rather high orbital inclination of at least
$i$ $\approx$ 49$^{\circ}$ (Moffat \& Seggewiss 1977;
Lenoir-Craig et al. 2021). A wide shock cone with a half
opening angle $\theta_{1/2}$ $\approx$ 90$^{\circ}$ would 
occur for nearly equal wind momenta $\mathcal{R}$ $\approx$ 1 
(Stevens et al. 1992) and large adjustments in the 
assumed mass-loss parameters would thus be required. 
But such a large opening angle is not consistent with the 
models of Hill et al. (2002) which gave
$\theta_{1/2}$ = 46$^{\circ}$$\pm$4$^{\circ}$.

The accuracy of the SB system orbital 
ephemeris is crucial for calculation of orbital phases
and interpreting line shifts. Determination of the SB orbital
parameters for $\theta$ Mus is difficult (Sec. 2).
The formal uncertainties in $P_{orb}$ and reference HJD 
for minimum light T$_{o}$ corresponding to $\phi$ = 0 adopted 
here (Table 1) are too small to produce a substantial error 
in the computed phases of the X-ray observations. Combining 
the uncertainties for $P_{orb}$ and T$_{o}$ results in a
phase error of  $\Delta\phi$ $\approx$ 0.08 (1.54 d) for
the June 2022 {\em Chandra} observations and only half
that amount for the {\em XMM-Newton} observations. Even so,
the X-ray phases are quite sensitive to the adopted period 
and a small error of 0.013 d in  $P_{orb}$ would shift the computed
{\em Chandra} phases by $\Delta\phi$ $\approx$ 0.25. 
If the shorter periods
obtained by Moffat \& Seggewiss (1977) or Marchenko et al. (1998)
were adopted the computed phases would change dramatically.
But the upper limit on orbital
period change $\dot{P}_{orb}$ $<$ $-$4.57 s yr$^{-1}$ 
derived by Lenoir-Craig et al. (2021) has a  negligible 
effect on computed X-ray phases. 
Additional X-ray observations with more complete phase
coverage over the 19 day orbit would be useful
to see if line shifts vary with phase or from red to blue.
There is already a hint for slightly larger redshifts
in the 2009 {\em XMM-Newton} RGS spectra than in 
2004 (Table 3). Also any future observations that
improve our knowledge of the $\theta$ Mus SB or
wide SB$+$OI system orbital parameters or the 
nature of the elusive WR companion (discussed by Hill et al. 2002)
might help clarify the X-ray results.

\section{Summary of Main Results}

New {\em Chandra} X-ray observations of $\theta$ Mus
confirm that emission lines in the cooler plasma component
are redshifted near $\phi$ $\approx$ 0.65 
(OV star passing in front of the WC star)
as previously reported from 2004 {\em XMM-Newton} observations.
An archived 2009 {\em XMM-Newton} observation
reveals that low-temperature 
lines are similarly redshifted at opposite phase 
$\phi$ $\approx$ 0.95. The persistent redshifts
could be attributed to a CW shock in the wide SB$+$OI 
system.  But blueshifted lines that should appear if
CW shock emission is also present in the short-period 
SB system have not been detected so far, for unknown reasons.
Assuming no large changes occurred in the intrinsic  
X-ray spectrum of $\theta$ Mus, the decrease in emergent hard-band flux
between the 2004 and 2009 {\em XMM-Newton} observations is
likely due to increased absorption toward the hotter plasma 
component. The high absorption inferred from spectral
fits points toward dense hot wind-shocked
plasma near the line-of-centers in the confines of the
SB system. The X-ray emission of $\theta$ Mus is quite similar to the
WC$+$O binary $\gamma^2$ Vel implying a common physical
mechanism and both systems show puzzling line centroid 
properties that are not easily reconciled with  CW models.

\clearpage

\begin{acknowledgments}
This work was supported by {\em Chandra} award GO1-22006X
issued by the Chandra X-ray Observatory Center (CXC). The CXC is operated by the
Smithsonian Astrophysical Observatory (SAO) for, and on behalf of,
the National Aeronautics Space Administration under contract NAS8-03060.
This work is partially based on  archival data obtained with {\em XMM-Newton},
an ESA sceince mission with instruments and contributions from
ESA and the USA (NASA).
This work has utilized HEASOFT developed and maintained by HEASARC at NASA GSFC
and the {\em XMM-Newton} SAS data analysis software package.

\noindent This paper employs a list of {\em Chandra} datasets obtained by the 
{\em Chandra} X-ray Observatory contained 
in~\dataset[DOI: 10.25574/cdc.173]{https://doi.org/10.25574/10.25574/cdc.173}.
\end{acknowledgments}

\vspace{5mm}
\facilities{{\em Chandra}; {\em XMM-Newton}}

\vspace{5mm}
\software{CIAO (Fruscione et al. 2006), HEAsoft (HEASARC 2014), XSPEC (Arnaud 1996), 
SAS (Gabriel et al. 2004)}

\end{document}